\documentclass[conference]{IEEEtran}
\IEEEoverridecommandlockouts
\usepackage{cite}
\usepackage{amsmath,amssymb,amsfonts}
\usepackage{algorithmic}
\usepackage{graphicx}
\usepackage{textcomp}
\usepackage{xcolor}
\def\BibTeX{{\rm B\kern-.05em{\sc i\kern-.025em b}\kern-.08em
    T\kern-.1667em\lower.7ex\hbox{E}\kern-.125emX}}
\begin{document}

\title{TrustChain: Trust Management in Blockchain and IoT supported Supply Chains\\
}

\author{Sidra~Malik,
        Volkan~Dedeoglu,
        Salil~S.~Kanhere,
        and~Raja~Jurdak
\thanks{S. Malik and S.S. Kanhere are with University of New South Wales, Sydney, NSW, 2052 Australia email: name.surname@unsw.edu.au}
\thanks{V. Dedeoglu and R. Jurdak are with CSIRO Data61, Pullenvale,
QLD, 4069 Australia e-mail: name.surname@csiro.au}
}

\maketitle

\begin{abstract}
Traceability and integrity are major challenges for the increasingly complex supply chains of today's world. Although blockchain technology has the potential to address these challenges through providing a tamper-proof audit trail of supply chain events and data associated with a product life-cycle, it does not solve the trust problem associated with the data itself. Reputation systems are an effective approach to solve this trust problem. However, current reputation systems are not suited to the blockchain based supply chain applications as they are based on limited observations, they lack granularity and automation, and their overhead has not been explored. In this work, we propose TrustChain, as a three-layered trust management framework which uses a consortium blockchain to track interactions among supply chain participants and to dynamically assign trust and reputation scores based on these interactions. The novelty of Trustchain stems from: (a) the reputation model that evaluates the quality of commodities, and the trustworthiness of entities based on multiple observations of supply chain events, (b) its support for reputation scores that separate between a supply chain participant and products, enabling the assignment of product-specific reputations for the same participant, (c) the use of smart contracts for transparent, efficient, secure, and automated calculation of reputation scores, and (d) its minimal overhead in terms of latency and throughput when compared to a simple blockchain based supply chain model. 
\end{abstract}


%
\IEEEpeerreviewmaketitle

\section{Introduction} \label{sec:introduction}
\begin{figure}[t!]	
\centerline{\includegraphics[width=9cm, height=7cm, keepaspectratio]{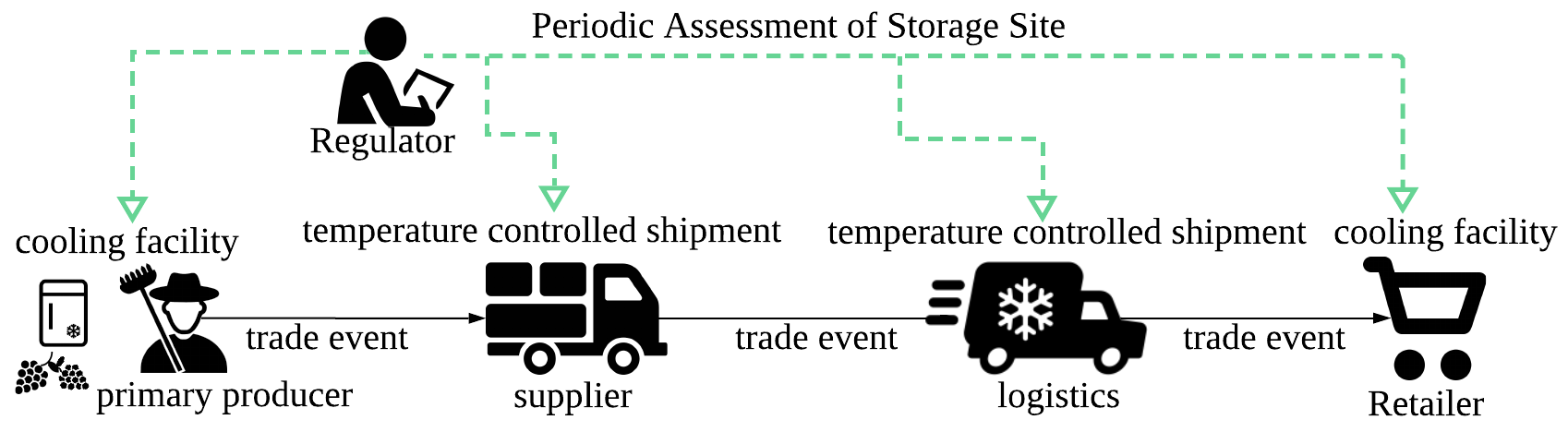}}
	\caption{A typical food supply chain may involve producers, suppliers, logistics, retailers, and regulators.}
	\label{fig:sc}
	\vspace{-0.4cm}
\end{figure}

\IEEEPARstart{B}lockchain (BC) is a time-stamped series of immutable data records which can enhance existing supply chains with traceability, provenance, ownership information and anti- counterfeiting. In BC-based supply chains, events such as trade, ownership and location data are hashed and linked to BC transactions. These transactions are grouped into blocks that are linked together with cryptographic hashes, making them immutable. The significance of the unsolicited confirmation of supply chain events  can be realised well in food supply chains where there is a need to trace the origin of products, or identify a point of fraud  
such as the horse meat scandal\cite{2013horse} or the source of an outbreak such as the salmonella virus in papayas \cite{papaya}.
In \cite{sidra2018productchain}, the authors showed that a consortium permissioned BC can make otherwise siloed supply chain event data available to all authorised participants making traceability more robust and time efficient. 
However, the integrity of data is identified as an unsolved problem for BC based supply chains. In this paper, we use food the supply chain as a representative supply chain application example as in \cite{sidra2018productchain} but note that the presented framework can be generalized to other supply chains. Figure \ref{fig:sc} depicts the supply chain of a food product starting from the primary producer to the retailer.


Most conventional BCs are based on the creation and transfer of digital assets. In these applications, BC provides not only immutability but also a proof that the stored data is correct and trusted. This feature is a result of the integration of creation and transfer of digital values with the distributed consensus mechanisms based on public key cryptography and digital signatures. For example, in Bitcoin BC, creation and transfer of bitcoins is integrated to the Proof of Work (POW) consensus mechanism. However, for physical commodity and asset trading applications, the hashed data on the BC represents digital observations of physical events. Although data related to supply chain events is immutable once recorded on a BC, the BC cannot ascertain the authenticity of observations provided by supply chain entities. The authenticity and trust of the data becomes questionable, and thus raises the concern of data integrity on the BC.\par

In this paper, we argue that despite BC being an effective technology for managing supply chain traceability, it alone cannot support the trust and reliability of data regarding the quality of commodities and the trustworthiness of supply chain entities. False data generated by the supply chain entities becomes immutable once recorded on the BC. One approach to improve the trust and reliability of the data is to use accountability and reward mechanisms to penalise and incentivise dishonest and trustworthy participants, respectively. These mechanisms rely on a trust management system (widely used in e-commerce, distributed systems) which we propose to integrate within a supply chain BC. Such a system may also benefit from the data generated by Internet of Thing (IoT) sensors (e.g.  temperature, location) which are being increasingly embedded in various stages of the supply chain life-cycle (e.g. farms, manufacturing plants, shipping containers) \cite{IBMfraud,prov}. However, IoT sensors are also susceptible to faults or malicious attacks and thus cannot be blindly trusted.  
Apart from IoT sensors, there are other observations contributing to trust in the real world, such as food authority approvals, the brand perception of a seller in a food market, etc. Moreover, current approaches for trust management either attribute the reputation to the entities (i.e., agent-based) or to the asset (i.e., resource-based)\cite{yang2017blockchain}. However, a supply chain entity (e.g. a supplier) may be simultaneously involved in the trade of multiple products, where reputation for each type of product becomes necessary. Supply chain applications demand more flexibility and granularity as we not only need to trust both the entities and the commodities, but also the entity within one particular product supply chain. Finally, the integration of the trust management system within the BC-based supply chain framework should introduce minimal overheads to the latency, throughput and resource consumption. 
In summary, some challenges to devise an effective reputation system in supply chains are: (a) 
the need for a multi-faceted assessment of the trustworthiness of the data logged in the BC which incorporates inputs from IoT sensors, feedback provided by supply chain entites, physical audits, etc. 
(b) a supply chain participant may trade more than one type of commodities; a participant must be evaluated distinctively for each of these types and so should be the individual commodity based on whether its quality was preserved during the product chain, (c) to action the penalties and incentives, an automated framework is required which not only provides traceability of supply chain events but also relates each of these events to a trust value of a participant and the quality of commodity, and (d) the associated overheads should be minimal and not impact the scalability of the platform. \par
To address the above challenges, we propose a three-layered BC-based trust management framework called TrustChain which makes the following novel contributions:
\begin{enumerate}

     \item A BC based reputation and trust framework for supply chains that operates at both agent and resource level and evaluates the truthfulness of data based on multiple sources of data. The framework offers flexibility to compute the reputation at different levels of abstractions ranging from the product to a particular supply chain entity and even the role of this entity in the supply chain of a specific product.
    
    \item 
    We leverage smart contracts for automation of reputation calculation with BC transactions and penalties to action the rewards and accountability for both supply chain participants and quality of food product being traded. Based on the output of the smart contracts, supply chain participants and commodities receive reputation scores as a measure of their trustworthiness for a trade event. Supply chain participants are then penalised by revoking their participation in the supply chain or rewarded by getting high ratings published.
    \item 
   We develop a complete implementation of TrustChain using Hyperledger Fabric. Our evaluations reveal that the mechanisms introduced by our framework add minimal overheads in terms of throughput and latency in comparison to a trading model that does not incorporate a trust management system on BC. We also perform a qualitative security analysis of TrustChain's resilience to known attacks against reputation systems. 
\end{enumerate}
The rest of the paper is structured as follows: Section \ref{sec:rw} includes  related  work followed by the TrustChain's framework in  Section \ref{sec:RepTrust}. Security analysis and performance evaluation is presented in Section \ref{sec:eval} and  conclusion in Section \ref{sec:conc}.

\section{Related Work}\label{sec:rw}
In \cite{khaqqi2018incorporating} the authors proposed a novel model for Emission Trading Scheme (ETS) which incorporates BC technology with reputation-based trading to address fraud and management issues in ETS. The system's efficacy is improved by the reputation-based market segmentation and priority-value-order mechanisms, which allow sellers with high reputations to access more and better offers from buyers, and also to sort offers based on reputation and price. However their approach is not granular as the sellers with different approaches to emission reduction schemes are always grouped together. Additionally, reputations of these sellers are drawn from only the observations of auditing bodies, which may not be as frequent as the trade transactions. As the reputations may not get updated frequently, this may favor the same group of sellers by always giving them access to higher bids, and may introduce bias in the selection of buyers. \par

 The authors in \cite{yang2017blockchain} propose a BC based reputation system to ensure the data credibility of vehicular systems.  A temporary center node among a cluster of vehicles is selected to generate ratings for nearby vehicles. Then, the vehicles in the cluster form a consensus for these vehicle ratings before they are stored on the BC. This results in a large number of messages associated with the proposed consensus mechanism to update the reputation of vehicles. The performance of the proposed BC based approach is compared to a scheme, where the ratings are verified by the vehicles by using message hashes  without a consensus. They analyse the performance of the system by introducing malicious vehicles and argue that using consensus for verifying the ratings is a better approach when there are malicious ratings generated by untrusted vehicles. However, the additional cost of the BC system, in terms of latency for reaching the consensus, is not evaluated.\par 
 Another promising approach for trustless, privacy preserving resource based reputation system is presented in \cite{schaub2016trustless}. Their approach highlights the benefits of anonymous rating of assets by preserving the privacy of consumers. However, without a performance analysis, it is hard to evaluate the overall competence of token generation, which is used by consumers to generate decoupled ratings from transactions. Also, in case of unfair ratings, some ratees would be disadvantaged as there is no direct link between a transaction and a rater which eventually can be exploited by malicious users.\par 
 
 In \cite{moinet2017blockchain}, a trust model is proposed in the context of autonomous wireless sensor networks, where the nodes must maintain a minimum trust level to keep participating in the network and avoid revocation. However, the proposed approach works on the network node level only and lacks granularity as a node may provide more than one service and must be evaluated for its trustworthiness accordingly. Moreover, the messages used for reputation calculations are verified by their message digests, which is the only criterion of message authentication.\par 
 
Recently, IBM\cite{IBMfraud} researchers have been developing crypto-anchors, tamper-proof digital fingerprints, which can be embedded into products and linked to a BC as a proof of product's identity. The input from crypto-anchors could be readily integrated in TrustChain and can potentially address the problem of counterfeiting, specially in pharmaceuticals and high-value assets. However, it would be interesting to note the cost analysis as it may be a limiting factor for the adoption of the technology in supply chains.\par

In summary,  the existing reputation systems for BCs are either asset-based or agent-based and fail to provide the level of granularity required by supply chain applications. Furthermore, the reputations are mostly sourced from single points of observations. There is also a lack of consideration of adversarial models and a quantitative analysis of the overheads associated with the introduction of the trust model. TrustChain is designed considering the aforementioned challenges and takes into account the multiple data observations, automation of rating calculations, accountability mechanisms, a detailed security analysis and performance analysis in terms of network throughput and latency.

\section{Reputation and Trust in BC based Supply Chains}\label{sec:RepTrust}
As discussed in Section \ref{sec:introduction}, traceability and integrity are major challenges faced by supply chain management systems. There are two basic requirements for providing traceability and integrity in supply chains: (1) the data that provides traceability and integrity of supply chain events, product data stating its properties, IoT sensor data and other supplementary sources of data (such as crypto-anchors) and regulatory endorsements should be recorded in a tamper-proof way, (2) the recorded data should be authentic and should represent the true observations of sensor devices, supply chain entities, and other data sources. The BC technology satisfies the first requirement with a distributed tamper-proof ledger. The aim of our work is to address the second requirement by devising mechanisms to establish trust in data at the point of origin and ensure that the data recorded on the BC is trusted. As supply chains involve multiple entities and product types, the trust should be established at a granular level that takes into account the different product types, the entities, and their interactions. Furthermore, the process can be automated providing real-time traceability. To achieve this aim, our proposed framework called TrustChain proposes a BC integrated trust and reputation module that evaluates the truthfulness of the supply chain data and calculates reputation scores for commodities and supply chain entities at a granular level. To automate the process, we employ smart contracts, which are self executing software programs invoked when predefined conditions are met.\par

\begin{figure}[!t]	\centerline{\includegraphics[width=3.5in, height= 3in,keepaspectratio]{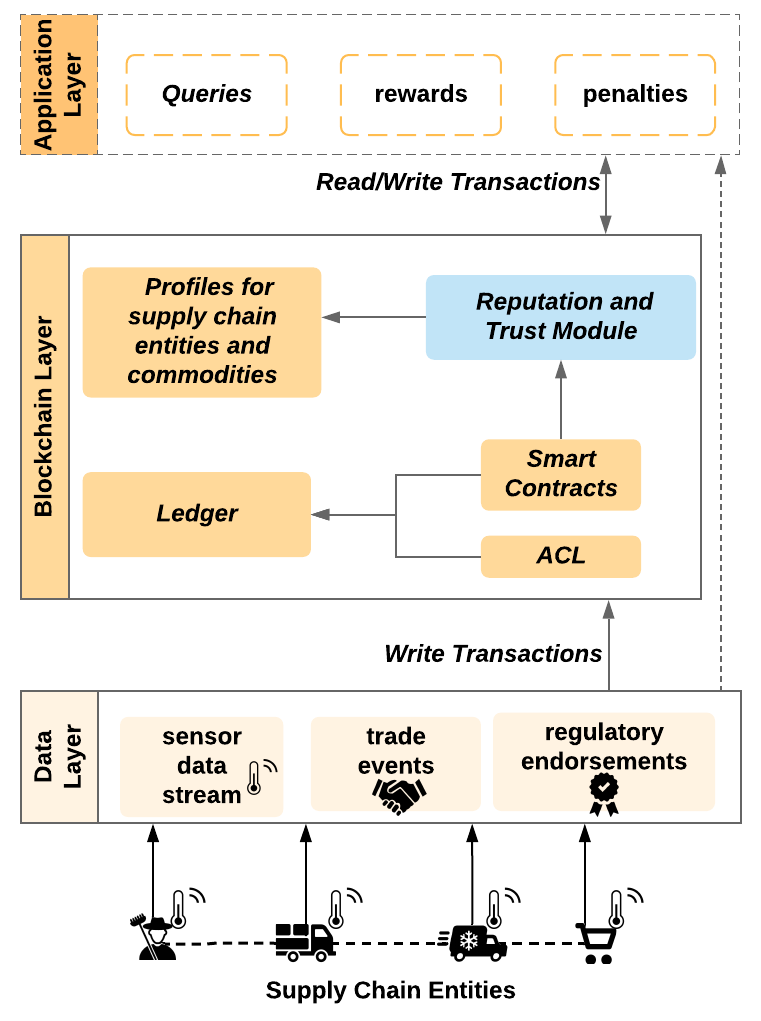}}
	\caption{Three-layered structure of the TrustChain framework}
	\label{fig:framework}
	\vspace{-0.5cm}
\end{figure}

 TrustChain framework is organized into three layers: data, BC and application with supporting components as shown in Figure \ref{fig:framework}. The data layer encompasses supply chain data produced by sensor devices, trade events between entities, and regulatory endorsements. The raw data can be stored in a database at the application layer (i.e., off-the-chain), while the message digest of the data is sent to the BC layer in the form of transactions. At the BC layer, the transactions are stored on the ledger and processed following a set of access rules defined by the Access Control List (ACL). The access rules specify who can read or write the data on the ledger. The transactions invoke smart contracts, which generate reputation and trust values for entities and quality ratings for commodities using the reputation and trust module. The smart contracts also emit warning events depending on predefined conditions (e.g., when a frozen product is stored above 0 degree). The reputation and trust values  are stored on the digital profiles of supply chain entities and commodities on the BC. Finally, the application layer interacts with the BC layer through queries. The administrators and regulators query about the trust and quality scores of entities and commodities respectively. The quality of commodity is also made available to the consumer when it finishes the product chain. Based on the retrieved scores, they action rewards and penalties, which recompense the entities with high scores by publishing their scores, penalizes the entities with low scores by revocation from the network, and publishes the product ratings for final consumers.\par

We implement TrustChain on a BC as a service platform, where the permissioned BC network is managed by a business network administrator sitting between the supply chain entities and the BC network. The business network administrator has administrative control over the BC and defines the business network model. We choose Hyperledger Fabric\cite{cachin2016architecture} for the deployment due to its support for business related applications, ease of deployment, and availability of tools for BC deployment, management, data query, smart contract implementation, and cross organisational collaboration. Note that, TrustChain can also be implemented on a Hyperledger Fabric based multi-organisation blockchain network\footnote{https://hyperledger.github.io/composer/v0.19/tutorials/deploy-to-fabric-multi-org} owned by a consortium of supply chain entities rather than using a BC as a service platform.\par
We assume that the entities maintain a static public key in order to be identified in the business network. When an entity submits a query/write transaction, the validating peers verify the transaction does not violate the ACL rules (see Figure \ref{fig:acl}).

In the following sections, we describe the layers of the TrustChain framework in detail.
\subsection{Data Layer}\label{sec:datapayloads}
As shown in  Figure \ref{fig:framework}, the data layer has inputs from sensor data streams, trade events, and regulatory endorsements. Based on these inputs, the data layer generates transactions for the BC layer. For scalability, the raw data collected at the data layer can be stored in a database (i.e. off-the-chain) rather than being stored in the BC. 

\textbf{Sensor data streams:} 
We assume that each supply chain entity from the primary producer to the retailer, as shown in Figure \ref{fig:framework}, has IoT sensors installed to monitor  temperature, location, humidity, etc. These sensor readings can be used as an indicator of the quality of the food products. In order to ensure the accuracy of reported readings, we further assume that these IoT sensors are calibrated periodically. In this paper,  
we use temperature sensors as an example. We assume that the commodity must be stored within a certain temperature range (max and min thresholds) at all times from its origin to the retailer's shelf. Based on the temperature readings, the commodity is given a rating, denoted by $Rep_{sens}(t)$. Moreover, warning messages can be generated if the reported sensor temperature is out of bounds of the desired range. The conditions for commodity rating and warning messages are specified in the smart contract related to the commodity (to be discussed in Section \ref{sec:smartcontracts}). \par

\textbf{Trade events:} In our previous work 
\cite{sidra2018productchain}, authors defined a trade event as a change of ownership of a commodity, which is stored on the BC. In Trustchain, in addition to the change of ownership, a rating, $Rep_{trader}(t)$, attributed by the buyer for the seller is appended to the BC. The rating is based on a quality assessment of the traded commodity, which is carried out on pre-negotiated terms between the trading parties, such as the use of crypto-anchors \cite{IBMfraud} or acceptance sampling\cite{starbird1997acceptance} at the buyer's end. These quality assessment conditions are specified in the smart contract (see Section \ref{sec:smartcontracts}). The assessment by the buyer incentivises the seller to trade honestly, and both trading parties would benefit from the honest trade. However, a buyer may intentionally generate a false rating for a seller which is termed in TrustChain as ``unfair rating" and dealt using a ``dissatisfaction flag" which can be issued by the seller. This is discussed further in Section \ref{sec:security}.\par

\textbf{Regulatory endorsements:} Regulatory endorsements are generated by food safety authorities performing physical on-site checks on production, storage/cooling, and transportation facilities, or the HACCP (Hazard Analysis and Critical Control Points) checks\cite{sidra2018productchain} to approve the quality conditions. Regulatory ratings, $Rep_{reg}(t)$, are generated based on the regulatory endorsement data in the forms of certificates and reports.\par 

All the ratings are generated using transactions and smart contracts explained in the next section. The generated ratings $Rep_{sens}(t)$, $Rep_{trader}(t)$, and $Rep_{reg}(t)$ at time $t$ are sent to the BC layer to be used by the trust and reputation module to calculate reputation and trust values for the supply chain entities and commodities.\par 
\subsection{BC Layer}\label{sec:BC comp}
\begin{figure}[t!]
	\centerline{\includegraphics[width=3.6in]{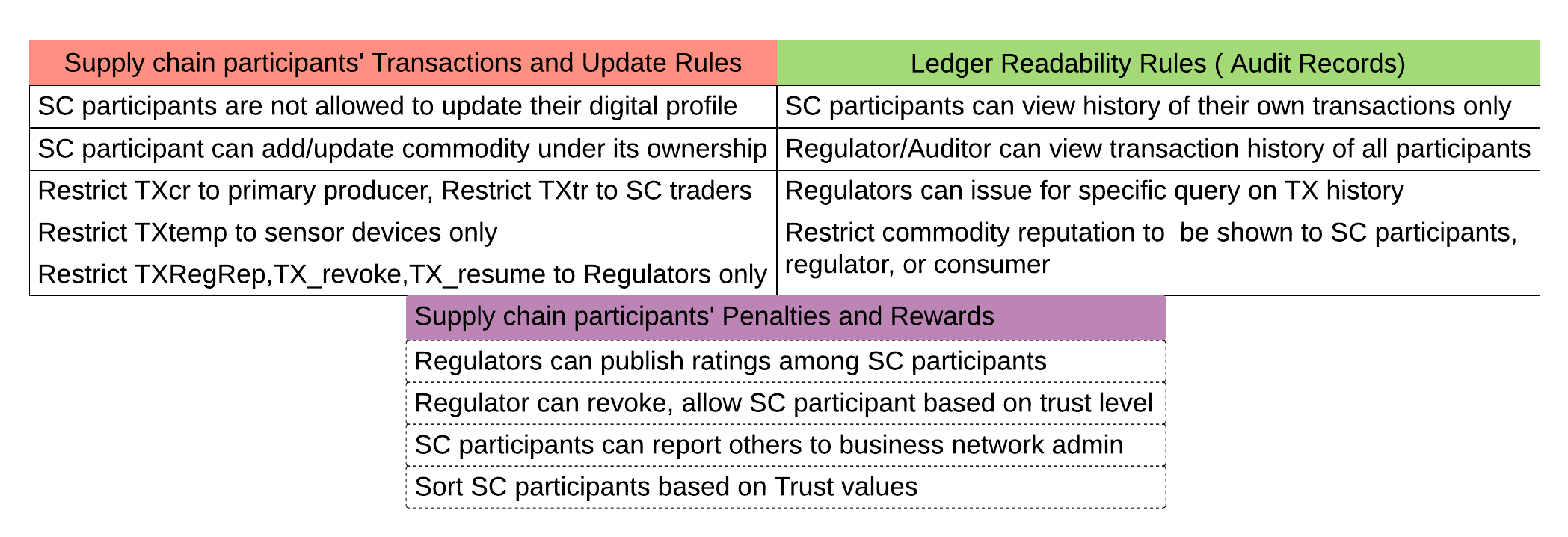}}
	\caption{ACL defines the rules of access to the resources of the TrustChain}
	\label{fig:acl}
	\vspace{-0.4cm}
\end{figure}

Transactions in TrustChain are governed by an ACL. Figure \ref{fig:acl} presents the TrustChain ACL, which  defines the permissions for submitting transactions, read/write access to the ledger, updating profiles, and other actions for the supply chain entities.  The BC layer maintains the digital profiles for all supply chain entities and commodities as shown in Figure \ref{fig:framework}. These digital profiles are mapped to the entities and the commodities using the enrolment ID of the certificates issued by a certification authority. TrustChain ACL does not give the entities the permission to modify their digital profiles (see Figure \ref{fig:acl}). Next, to automate the reputation calculation, smart contracts are invoked by transactions received from the data layer. 
The following sections present the transactions, the smart contracts, and the reputation and trust module of the BC layer in detail.
    \subsubsection{Transactions} \label{sec:txns}
    We adopt our transaction vocabulary from\cite{sidra2018productchain} and categorize the transactions based on the layer invoking them. This section describes the transactions invoked at the data layer which records supply chain events (eg. existence of a new commodity), change in the state of the entities (eg. reputation of a trader) and the commodities (eg. change of ownership) on the ledger. 
    Transactions at the data layer include the create transaction ($TX_{cr}$), the trade transaction ($TX_{tr}$), the sensory transaction ($TX_{sens}$), the regulator transaction ($TX_{Reg})$, and the receipt of commodity transaction ($TX_{rec}$) with their details as follows:\par
    A ledger that stores information pertaining to any particular commodity starts with a create transaction $TX_{cr}$, which is issued by a primary producer to confirm the existence of a new commodity and identify the quality smart contract this commodity will be bound to. A quality contract specifies pre-negotiated terms and conditions regarding the quality assessment, max/min temperature bounds, rating criterion etc. 
    The $TX_{cr}$ is given by:
    \begin{equation}
    TX_{cr}= [CID|H_{data}|ID_{o}|ID_{contract}|Sig_{o}|PU_{o}]   
    \end{equation}
    where $CID$ is the identifier of the commodity, $H_{data}$ is the hash of the commodity data (e.g. commodity type, quantity, unit price, etc.), $ID_{o}$ is the commodity owner's identifier (i.e. the primary producer), and $ID_{contract}$ is the identifier of quality contract the commodity is bound to. $Sig_{o}$ and $PU_{o}$ are the signature and the public key of the commodity owner.\par  
    After a commodity is created by the primary producer, it can be traded between different supply chain entities as it makes it way to the retail shelf (see Figure \ref{fig:sc}). The trade transaction confirms the physical handover of a commodity from the seller to the buyer:
    \begin{equation}
    TX_{tr}= [CID|H_{data}|ID_{b}|Sig_s|PU_s|Sig_b|PU_b]
    \end{equation}
    where $CID$ and $H_{data}$ are similar to $TX_{cr}$, and $ID_{o}$ is replaced with $ID_{b}$, the identifier of the buyer. $Sig_s$, $PU_s$, $Sig_b$, and $PU_b$ are the signatures and the public keys of the seller, and the buyer, respectively.\par
    As the commodity gets registered using $TX_{cr}$, IoT devices monitoring the temperature information of a commodity can log the temperature related data on the BC using sensory transactions, $TX_{sens}$. These transactions are generated by the gateway nodes identified by device IDs as IoT sensors have typically low computational capabilities. It is important to note that the generation rate of this transaction is decoupled from the rate at which the commodity is actually traded. This is to ensure that the commodity is being monitored at regular intervals while it is in storage, prior to a trade event. A sensory transaction is given by:

    \begin{equation}
    TX_{sens} = [CID|H_{data}|Sig_{device}]
    \end{equation}
    where $CID$ is the identifier of the commodity, $H_{data}$ is the hash of the IoT sensor stream, and $Sig_{device}$ is the signature of the gateway node generating the sensory transactions.\par
   The regulator in TrustChain is assumed to be a regional authority enforcing food safety standards such as Food Standards Australia New Zealand (FSANZ) \cite{fsanz}. As described in Section \ref{sec:datapayloads}, after a physical inspection of a storage site, the regulator issues a rating for the seller, $Rep_{reg}(t)$ via the regulator transaction $TX_{Reg}$ given by:
     \begin{equation}
   TX_{Reg} = [ID_s|H_{data}|C_{type}]
 \end{equation}
 where $ID_{s}$ is the seller's identifier, $H_{data}$ is the hash of the inspection evidence, and $C_{type}$ specifies the type of commodity for which the score is issued. This ensures that the regulator's ratings are distinctively recorded for each type of commodity as the storage conditions and assessing bodies may differ for different types of commodity trade. Also, to ensure compliance with the quality and safety standards, it is important for the regulator to perform periodic checks on the site and issue updated ratings. Note that, these on-site inspections are not frequent. If the site has not been inspected by the regulator after a certain time period, i.e. the regulator's rating has an older timestamp than a certain inspection period, the weight of the regulator's rating is reduced so that the rating and trust module fairly gives more weight to other observations which are more current. \par
 
 Upon receipt of a commodity at the retailer's end, the receipt of commodity transaction $TX_{rec}$ is generated  to log the end of product chain on BC. 
  \begin{equation}
   TX_{rec} = [CID|Sig_r|PU_r]
 \end{equation}
 where $CID$ is the identifier of the received commodity at the retailer, $Sig_r$ and $PU_r$ are the signature and the public key of the receiving retailer. The purpose of introducing this transaction is twofold: (1) for security purposes, it is important to keep track of commodities completing the product chain as the commodities with no progressive chains may be indicated as fake, and (2) it invokes the quality smart contract (See Section \ref{sec:qc}) and updates the overall rating $Rep_{sens}(t)$ of the commodity throughout the product chain which can be made visible to the consumers via the application layer.
\begin{figure*}[h]	\centerline{\includegraphics[width=5.5in]{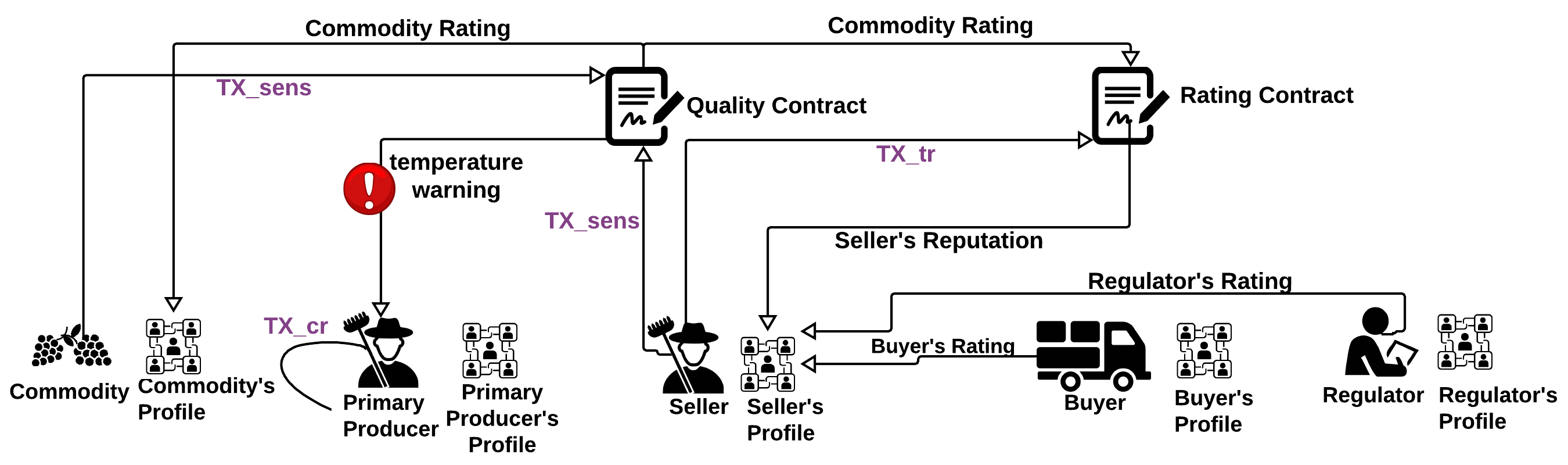}}
	\caption{Reputation calculation of seller using quality and rating contract}
	\label{fig:scContract}
	\vspace{-0.4cm}
\end{figure*}
    \subsubsection{Smart Contracts} \label{sec:smartcontracts} As discussed in Section \ref{sec:txns}, we use smart contracts in TrustChain for the calculation of ratings for entities and commodities when transactions pertaining to a supply chain event are recorded on BC. Using smart contracts, these ratings are calculated in a transparent, secure, efficient, and automated way, eliminating the need for any intermediaries. Together with the rewards and penalties mechanisms at the application layer, the ratings calculated by the smart contracts incentivise supply chain entities to contribute only trustworthy data to the network. 
   
    The smart contracts in Hyperledger are installed by the business network administrators or a subgroup of an organization. In case of multiple organizations, Message Service Provider (particular to Hyperledger) defines the local access rights for installing the smart contracts. More information on administrative access can be found in \cite{cachin2016architecture}. 
    Figure \ref{fig:scContract} depicts how these contracts contribute to the ratings of the commodities and traders with respect to a trade event on the supply chain. \par
    
   \textit{ \textbf{Quality Contract}:} \label {sec:qc}
   A quality contract is installed on TrustChain for every supply chain commodity. The contract enlists the quality rating criteria, i.e. temperature thresholds for the commodity, along with the contract identifier, $ID{contract}$. The quality contract categorizes temperature thresholds as: (1) boundary thresholds (the maximum upper and lower bounds of the required temperature in which the commodity is considered safe), 
   (2) damage thresholds (temperature thresholds exceeding which results in complete spoilage of the commodity).\par
   The quality contract is instantiated when a primary producer issues $TX_{cr}$, which binds the created commodity to the quality smart contract. The inputs to the quality contract are the temperature readings submitted by $TX_{sens}$. 
   Based on the temperature inputs and the temperature bounds defined in the contract, the quality contract generates two types of outputs: the warning notifications, and the reputation score of the commodity. The warning notifications are alerts for the commodity owner and issued when the temperature reading in $TX_{sens}$ reaches the  damage thresholds. The reputation score of the commodity $Rep_{sens}(t)$, is invoked as a result of a trade transaction, $TX_{tr}$, and is used by the rating contract to compute the reputation of the seller as shown in Figure \ref{fig:scContract}.\par

   The quality contract generates $Rep_{sens}(t)$ based on the temperature history of the commodity. If the temperature bounds defined in the contract are not violated, a maximum score is assigned to the commodity, whereas when the temperature readings indicate a damage, the respective score is reduced. The reputation score is updated with each trade event as the commodity moves through the product chain till it reaches the final retailer. When a commodity's product chain has $n$ trade transactions, we store the corresponding commodity reputation scores in the commodity's profile by generating $Rep_{sens}$:   
  \begin{equation}
  Rep_{sens}= [Rep_{sens}(t_0), Rep_{sens}(t_1), ..... , Rep_{sens}(t_{n-1}) ]
  \end{equation}
 
  When the commodity reaches the final retailer, the retailer generates $TX_{rec}$, which invokes the quality contract to generate the overall commodity rating, $R_{sens}$. This quality rating can be encoded on a QR code to be used by the retailer.\par
  
\textit{\textbf{Rating Contract}:} \label{sec:rc}For a trade event between a buyer and a seller happening at time $t$, the rating contract is invoked to compute the reputation of the seller $Rep_{seller}(t)$ from three inputs: (1) reputation score of the traded commodity $Rep_{sens}(t)$, (2) the regulator's rating for the seller $Rep_{reg}(t)$, and (3) the buyer's rating for the seller $Rep_{trader}(t)$. $Rep_{seller}(t)$ can be computed as a weighted sum:
\begin{multline}
  Rep_{seller} = w_1 \times Rep_{sens}(t) + w_2 \times Rep_{trader}(t) + \\ w_3 \times Rep_{reg}(t)   
\end{multline}
where $w_1$, $w_2$, and $w_3$ are the weighting factors for the reputation components. The choice of the weighting factors depends on the supply chain characteristics. For example, if the sensor data is infrequent, less weight can be given to $Rep_{sens}(t)$. The values of the weighting factors are chosen by the consortium members, and implemented by the business network administrator during network initialization. Moreover, the weights can be adjusted based on the available information (e.g., if a regulator rating is outdated, $w_3$ can be reduced). The overall reputation scores of a seller for each type of commodity are aggregated over a time period using the reputation and trust module described in the next section.

\subsubsection{Reputation and Trust Module}\label{sec:repmodel}

Once the reputation scores are calculated by the smart contracts, a reputation model can be chosen based on an aggregation function (mean, median, or beta-reputation \cite{commerce2002beta}). For TrustChain, we consider the reputation and trust management proposed by \cite{moinet2017blockchain}. Based on Gambetta's definition of trust \cite{gambetta2000can}, we adopt a time-varying and amnesic trust score calculation that adapts to supply chain events, where the recent events are given higher weights than the older events.

When a trader joins the supply chain network for the first time at time $t'$, it has no past reputation to compute the trust score $T_{trader}(t')$. Thus, the initial trust score assigned to the trader is $T_{trader}(t')= Trust_{min}$, which is the minimum trust score that each entity should maintain to keep participating in the TrustChain network. After initialization, the trust score of the trader is updated by the reputation and trust module. The calculation of the trust score involves two steps: (1) calculating the overall reputation score of the trader based on the current and previous reputation scores, and (2) calculating the trust score based on the overall reputation score and other application-specific features.

To calculate the overall reputation score $R(t_n)$ for a trader at time $t_n$, we consider the reputation scores corresponding to the current and previous supply chain events $Rep_{seller}(t_0)$, $Rep_{seller}(t_1)$, $, . . . . ,$ $Rep_{seller}(t_n)$ as:
\begin{equation}
    R(t_n) = \sum_{t={t_0}}^{t={t_{n}}} Rep_{seller}(t) \times \beta (t_n - t)
\end{equation}
where $Rep_{seller}(t_0)$ is the initial reputation score of the trader, and $Rep_{seller}(t)$ is the reputation score of the trader corresponding to the supply chain event happening at time $t$ with the corresponding forgetting factor $\beta(t_n-t)$. The forgetting factor $\beta(t)$ is a decaying function (e.g. $\beta(t) = e^{-f(t)}$ ), which enables the overall reputation score to evolve in time. Thus, the effect of the recent events in supply chain dominates the events happened earlier in time. Note that $R(t_n)$ is the overall reputation of a trader for trading a single type of commodity. For a trader trading multiple types of commodities, this score is calculated for each type of commodity, and stored in the profile of the trader separately to provide granularity. The reputation of the trader can be calculated on a periodic basis, where the period is defined by the network business administrator or the consortium of administrators defining the business model.

\subsubsection{Trust Evaluation}
Recall from the previous section that every trader should maintain the minimum trust level $Trust_{min}$ to participate in TrustChain. However, this trust value must be updated with increasing/decreasing reputation of a seller. Also, apart from the reputation score based on the ratings of buyers, commodities, and regulators, there may be other application-specific features affecting the trust score of a trader (e.g., number/volume of sales, consumer feedback for a retailer or commodity, etc.). Thus, we calculate the trust score of a trader $T_{trader}(t_n)$ based on the overall reputation $R(t_n)$ and some other feature scores $f_1, f_2, ..., f_N$ as:
\begin{equation}
    T_{trader}(t_n) = \alpha_0. R(t_n) + \alpha_1.f_1 + \alpha_2.f_2 +... + \alpha_N.f_N
\end{equation} 
where $\alpha_0$, $\alpha_1$,..., and $\alpha_N$ are the weighting factors determined by the business network administrator. As an example, let us consider the number of successful transactions a trader has completed as a feature to be used in the trust score. As an illustrative example, the business network administrator can assign the score for this feature shown by the Table \ref{tab:trust}. Similar to the reputation score calculation, a forgetting factor can be applied for the other feature scores.
\begin{table}[h]
    \caption{Trust Feature and Feature Score}
    \label{tab:trust}
    \centering
    \begin{tabular}{|c|c|lll}
    \cline{1-2}
    \multicolumn{1}{|l|}{\textbf{\begin{tabular}[c]{@{}l@{}}Number of Successful\\  Transactions\end{tabular}}} & \multicolumn{1}{l|}{\textbf{Feature Score ($f_1$)}} &  &  &  \\ \cline{1-2}
    0                                                                                                      & -1                                    &  &  &  \\ \cline{1-2}
    1-3                                                                                                      & 0.5                                    &  &  &  \\ \cline{1-2}
    4-6                                                                                                      & 1.5                                     &  &  &  \\ \cline{1-2}
    \textgreater{}= 6                                                                                        & 2                                     &  &  &  \\ \cline{1-2}
    \end{tabular}
\end{table}

The trust scores calculated by the trust and reputation module are stored in the profiles of the traders. When $T_{trader}(t_n)$ \textless  $Trust_{min}$, a notification message is generated to inform the network administrator about the trust level violation. Upon receiving the notification, the network administrator can revoke the trader from participating in the network. 

\subsection{Application Layer:}\label{sec:app}
The main function of the application layer is to address queries and transaction requests from administrators, regulators and consumers. Based on the query results and the role of the issuing authority, penalties and rewards are implemented by the application layer. 
\subsubsection{Queries}
The queries issued at the application layer include transactions to read/write and read only from BC. The read/write query transactions read the data from the ledger first, and then perform some computations on that data. For example, $T_{trader}(t_n)$ is computed and stored in the seller's profile, after reading the reputation $R(t_n)$ of the seller and other features described in the reputation and trust module. Note that, with every $TX_{tr}$, the rating smart contract automatically computes $Rep_{seller}(t_n)$. However, it would be cumbersome to compute $R(t_n)$ for every seller due to the significant delay it would cause on the validating peer's end. 
The network administrator can issue a transaction to request the reputation and trust module to calculate the overall reputation of a trader $R(t_n)$ over a certain period of time. Note that when $R(t_n)$ is calculated, the respective trust score, $T_{trader}(t_n)$, is also computed by the TrustChain implementation and the values are stored in the profiles of the entities.\par

The read query transactions, on the other hand, read data from the ledger and return the results to the application layer. For example, an end consumer can request the rating generated from sensor data stream $Rep_{sens}(t)$ for a commodity through its product chain. Other read query transactions can be generated for the properties of the commodities and the traders i.e. the traders with highest number of trades, the transactions with complete or incomplete product chains, commodities with high reputation scores etc. \par

\subsubsection{Rewards}
Supply chain entities, which trade honestly, should be incentivised to keep trading honestly. Thus, we introduce ``publishing the entities with highest trust values on the network" as a reward mechanism \cite{khaqqi2018incorporating}. Network administrators can publish the entities with high trust scores periodically, which would help the honest entities to find new customers and increase their sales. 

\subsubsection{Penalties}
The trust scores can also be used to penalize the entities with low scores, which do not trade honestly. The network administrator can penalise entities by revoking the entities from participating in the network for a certain period of time by invoking the $TX_{revoke}$ transaction, which updates the participation status of the entities in their profiles. Also, the network administrator can publish the list of revoked participants on the network for discouraging the other supply chain entities to contribute dishonestly. After the penalty period, the administrator resumes the participation of the entities with the $TX_{resume}$ transaction, which reverses back the the participation status of the entities. $TX_{revoke}$ and $TX_{resume}$ transactions are the only write transactions on the application layer.

\section{Evaluation and Results}\label{sec:eval}
In this section we provide a qualitative security and privacy analysis, a proof of concept implementation and a quantitative performance evaluation of the TrustChain framework. 
\subsection{Security and Privacy Analysis} \label{sec:security}
In  this  section,  we  discuss the known attacks in reputation systems and TrustChain's ability to defend against them. Regulators, business network administrators, and Hyperledger peers are considered to be honest, and excluded from  the threat model of TrustChain. Hyperledger endorsing peers are excluded from threat model as any transaction maliciously endorsed by one peer eventually gets checked by all the validating peers before it is committed to the ledger. Thus, due to BC's endorsement policies and consensus mechanisms, It is highly unlikely for the malicious peer to get successful.
Recall that TrustChain aims to solve the supply chain trust problem associated with the quality of commodities and the entities logging data on the blockchain. Thus, adversaries in our threat model include supply chain entities who are able to, individually or in collusion with the other supply chain entities, fake the data source by: modifying sensor feeds, creating  false  commodities, registering with multiple IDs, generating false ratings for other supply chain entities, masquerading as another identity, and not acknowledging of trade events; all to earn high ratings on the network dishonestly.   
Table \ref{tab:attacks}  summarizes  these attacks and explains various  mechanisms incorporated in TrustChain to protect against them. Note that, in Table~\ref{tab:attacks} we only consider the attacks against reputation systems and exclude network attacks in general. Based on the European Telecommunications Standards Institute (ETSI) risk analysis criteria~\cite{etsi2011102}, we also analyse the likelihood of the occurrence of these attacks and TrustChain's resilience against them.

TrustChain exhibits  high  resilience against eight out of the nine attacks, and moderate resilience against the remaining attack considered in Table~\ref{tab:attacks}. 
 We briefly explain here the problem of unfair ratings (see section \ref{sec:txns}), for which the seller can issue a ``dissatisfaction flag" for the dishonest buyer based on the proof needed to evaluate the quality. The flag message is sent to a validator who recalculates the $Rep_{seller}$. However, this could result in an incessant loop if the seller always issues a dissatisfaction flag for the honest buyer deliberately. The validator can resolve this by reducing the weight of the buyer's rating $w_2$, and re-evaluating $Rep_{seller}$ if: (i) multiple sellers have raised dissatisfaction flags for the same buyer, and (ii) the number of the consecutive dissatisfaction flags raised by the seller for the buyer is less than the number of trade transactions between them. These conditions indicate that the seller's dissatisfaction flags are genuine.\par

\begin{table*}[htbp]
\caption{Security Attacks}
\label{tab:attacks}
\scalebox{0.8}{
\begin{tabular}{|p{2.5cm}|p{7cm}|p{8cm}|p{1.5cm}|p{1.5cm}|}
\hline
\textbf{Attack} & \textbf{\textit{Description}}& \textbf{\textit{Defence}} & \textbf{\textit{Attack Likelihood}}& \textbf{\textit{Resistance to Attack}}\\
\hline
Sensor Tampering &	An attacker/ dishonest trader tries to tamper the sensors so that the sensors provide acceptable readings&	We rely on the existing methods as provided in \cite{diallo2018trustworthy},\cite{brodsky2018overlapping},\cite{saberi2018blockchain} for tamper proofing the sensors&	Possible &	High\\
\hline
Sensor feed modification &	An attacker tries to modify the sensor feed during communications &	The sensor data can be encrypted with shared key between the sensor device and the gateway. &	Possible& 	High \\
\hline

Whitewashing attack &	A dishonest trader tries to reset his poor trust score by re-joining to the system with a new identity and obtaining $Trust_{min}$. &	a)	The traders are registered to the permissioned network through a central authority. 
b)	In case of a revoked identity,  traders can re-join only if the network administrator allows. &Unlikely&	High	
\\
\hline
Sybil attack &
An attacker creates multiple identities (sybils) and tries to exploit them in order to manipulate a reputation score.&	Trader participants cannot create multiple identities& Unlikely & High
\\
\hline
Ballot Stuffing attack: Case1
&	A dishonest trader tries to generate false transactions with himself in order to give high reputation to himself & A trader is restricted from trading with himself according to access control restrictions &	Unlikely &	High \\
\hline
Ballot Stuffing attack: Case2 &A dishonest trader colludes with other traders in order to give high reputation score to himself by creating non-existent commodities &	a)	Every trade is bound to a physical commodity, and the presence of an invalid commodity can be detected as the product chain will not be completed when the commodity is not traded 
b)	The auditor of the business network can detect a trader with multiple obsolete genesis transactions and revoke the trader or lower his reputation score &	Possible &High\\
\hline
Bad Mouthing attack	&A trader, alone or colluding with other traders, tries to give a negative feedback to an honest trader in order to lower his reputation. &	Our system includes a dissatisfaction flag for the buyer in case the seller identifies a dishonest buyer. The dissatisfaction flag associated with a trader can be monitored by the validators to identify dishonest traders. &Possible&	Moderate\\
\hline
Impersonation and reputation theft&	A trader tries to acquire the identity of another trader and steal his reputation.	&Network registrations are refreshed by the business network admin after a certain time which may require proof of identification.	&Unlikely &High	\\
\hline
Repudiation of transactions &	A trader may deny a supply chain event that has happened or supply chain data that has been produced &	Transactions on BC ensure the immutability of data and the existence of trade events as well as the source of the data&	Unlikely& 	High\\
\hline
\end{tabular}
}
\label{tab:IP}
\end{table*}
\subsection{Proof of Concept Implementation}\label{sec:poc} A proof of concept implementation of TrustChain has been developed using Hyperledger Composer\footnote{https://www.hyperledger.org/projects/composer}, which is a high-level tool that facilitates building and running applications on top of the Hyperledger Fabric platform. For the implementation, we have modelled a business network on Hyperledger Composer for commodity trading among three supply chain entities: primary producer, shipper and a retailer following the transactional vocabulary described in Section \ref{sec:RepTrust}. To submit transactions and queries to the deployed blockchain business network, we have used the REST APIs. The end-to-end commodity trade in the TrustChain implementation is 
described by the folowing steps: 
\begin{enumerate}
    \item An instance of the quality contract is created for the commodity with the  temperature conditions specific to the type of the commodity.  
    \item The supply chain entities register on the network and are assigned initial trust scores $Trust_{min}$.
    \item As discussed in Section \ref{sec:txns}, when the commodity is ready for trade, the primary producer generates the $TX_{cr}$ which is bound to the ID of the quality smart contract $ID_{contract}$, and also assigns $CID$ to the commodity. 
    \item Once the $TX_{cr}$ is committed in the ledger, $TX_{sens}$ transactions indicating the temperature conditions of the  commodity are stored on BC and alerts are generated if the temperature readings exceed the thresholds defined in the contract. 
    \item The primary producer's storage facility is periodically assessed by the regulator and the respective $Rep_{reg}(t)$ is stored in the primary producer's profile. 
    \item The transaction $TX_{tr}$ is created for the commodity trade between the primary producer and the shipper, who becomes the new owner of the commodity. The shipper gives the rating $Rep_{trader}(t)$ to the primary producer based on the quality of the received commodity.   
    \item The transaction $TX_{tr}$ triggers the rating smart contract which computes $Rep_{sens}(t)$ and the primary producer's rating $R_{seller}$, and stores them in the commodity's and the primary producer's profiles, respectively.
    \item The steps 5-7 are repeated for the trade between the shipper and the retailer by replacing the primary producer and the shipper with the shipper and the retailer, respectively. The retailer is the final buyer of the commodity.
    \item Finally, the retailer issues $TX_{rec}$ which also generates the commodity rating using the quality smart contract. 
    \item Steps 1-9 are repeated for other commodities of the same type for a trade event happening at time $t$, and a seller has $[Rep_{seller}(t_n), Rep_{seller}(t_{n-1}), . . . , Rep_{seller}(t_0)]$ for each commodity type. The regulators and the administrators can request to compute $R(t_n)$ and $T_{trader}(t_n)$ which will be stored in seller's profile. 
 \item A seller is either penalised or rewarded based on the associated trust values. Consumers of the commodities can query $Rep_{sens}$, to check the temperature threshold violations throughout the product chain. Other queries regarding provenance, product chain and the seller's reputation can be invoked by administrators and regulators. 
\end{enumerate}

\subsection{Performance Evaluation}
The proof of concept implementation in Section \ref{sec:poc} is evaluated using Caliper\footnote{https://www.hyperledger.org/projects/caliper}, which is a benchmark tool that evaluates the Hyperledger performance. It allows Hyperledger users to measure the performance of a BC model using parameters such as latency and throughput. However, Caliper has a limited set of predefined network models to choose from. For TrustChain, we consider a base model of a solo orderer node and the two endorsing peers from separate organizations with one communication channel. The complete business network is modeled using Hyperledger Composer and performance tests are carried out using Caliper tool on a Dell Notebook (Intel Core i7, 2.21 GHz, 8 GB memory). \par 

For the performance evaluations, we consider the $Tx_{cr}$ and $Tx_{tr}$ transactions in TrustChain as they are not only frequent but also their computational overhead is higher. We consider a baseline BC system, without trust management, which only stores ownership information of supply chain events.  We then provide a performance comparison of the TrustChain with the baseline BC system by considering the $Tx_{tr}$ transactions only. To compute the overhead of our trust management system, we use $Tx_{tr}$ transactions, as these transactions invoke the rating smart contracts for the trust management system.\par

Next, we present the throughput and latency performance evaluations, where we vary the transaction send rate of $Tx_{cr}$ and $Tx_{tr}$ from 10 to 100 transactions per second (tps) for a simulation interval of 100 seconds. The evaluation results are obtained by averaging 10 runs per send rate for each type of transaction.\par

\subsection{Throughput Performance of TrustChain }
\label{tput}
Throughput is defined as the rate at which transactions are committed to the ledger after being issued from a trader.\par
\begin{figure}[t]
\centerline{\includegraphics[width=2.9in]{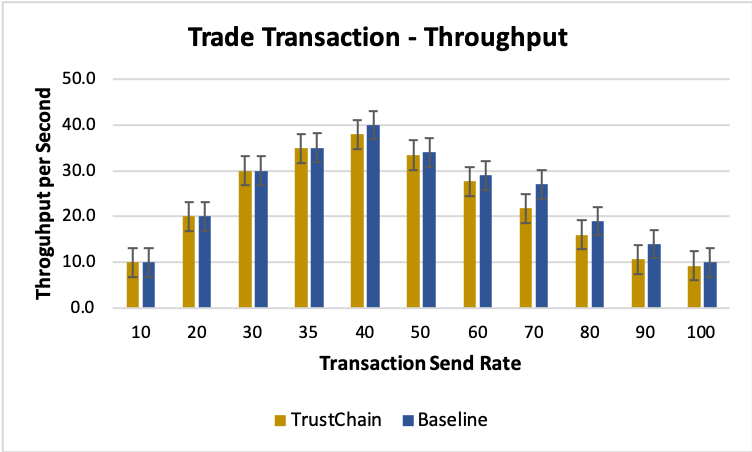}}
	\caption{Comparison of average throughput}
	\label{fig:tradet}
	\vspace{-0.4cm}
\end{figure}

 
\textit{Trade transactions:} $Tx_{tr}$ is the most expensive transaction in TrustChain as it involves computing the $Rep_{seller}$ along with updating the ownership of a commodity.
The throughput comparison of $Tx_{tr}$ in baseline BC and TrustChain are shown in Figure~\ref{fig:tradet}. We observe that the additional overhead introduced by the Trustchain is in the order of a few seconds, resulting in the throughput of the baseline system being higher by only 5 transactions at 70tps.
Figure~\ref{fig:tradet} also shows that the throughput increases linearly as expected till it reaches the maximum at around 40 tps and starts decreasing beyond this point. This represents a saturation point for a validating peer to handle the increasing transaction rate. This trend is similar to a congestive collapse in network congestion, when after reaching a saturation point the network settles into a state where it generates very low throughput, high packet loss and an exponential increase in packet delays. The similar trend is noted in the baseline system, where the throughput starts decreasing around 40-50 tps.\par

\textit{Create transactions:} Recall from Section \ref{sec:BC comp}, that with each $TX_{cr}$, a new commodity is generated. Whereas, $Tx_{tr}$ is updating the status of already generated commodities and reputation scores of participants. Thus, it excludes the validation of new resources when an item is registered on the network. For $TX_{cr}$, Figure \ref{fig:createtl} shows that the increasing transaction send rate  throttles the validators around 35-40 tps, slightly earlier than that in $Tx_{tr}$.  This is most likely due to the additional delay added by resource validation in $Tx_{cr}$, resulting in high latency for $Tx_{cr}$ which is detailed in the next section.

\subsection{Latency Performance of TrustChain}
\label{lat}
\begin{figure}[t]
\centerline{\includegraphics[width=2.9in]{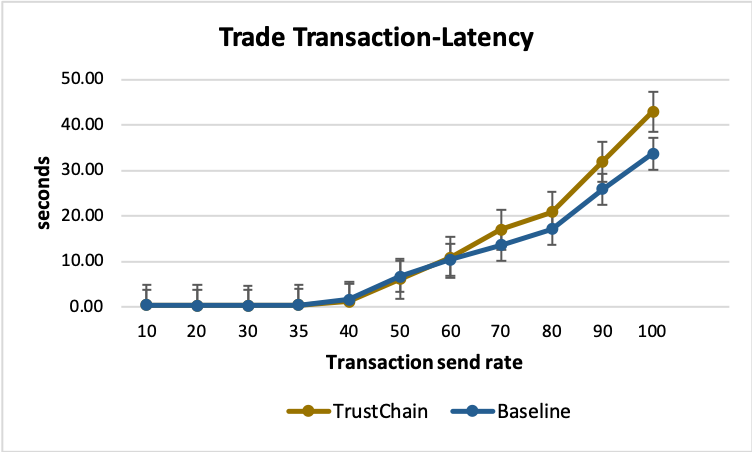}}
	\caption{Comparison of average latency}
	\label{fig:tradel}
\end{figure}

\begin{figure}[t]
\centerline{\includegraphics[width=3in]{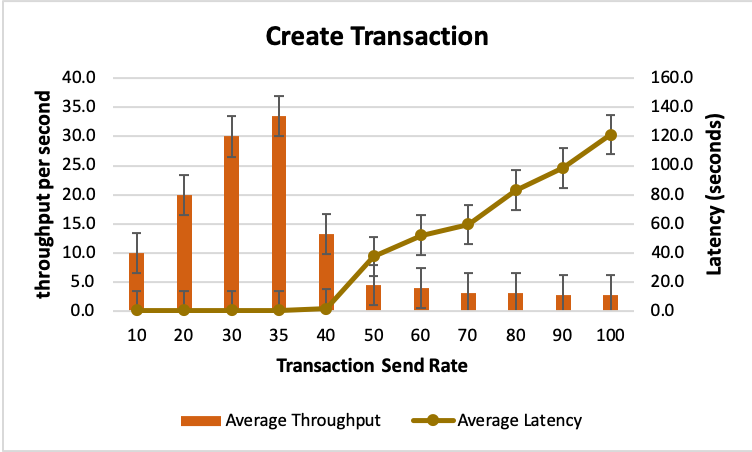}}
	\caption{Average throughput and latency}
	\label{fig:createtl}
	\vspace{-0.4cm}
\end{figure}
Latency is the time taken from an application sending the transaction to the time it is committed to the ledger by the committing peer in Hyperledger. 
From \cite{thakkar2018performance}, we deduce that the transaction latency comprises of endorsement latency, broadcast latency, commit latency and ordering latency. Caliper however provides only the cumulative delay and not a detailed breakdown of the aforementioned components. \par
Figure \ref{fig:tradel} and \ref{fig:createtl} show the average latency for $Tx_{tr}$ and $Tx_{cr}$, respectively. It is noted from Figure \ref{fig:tradel}, that the average latency of TrustChain is slightly higher than the baseline. Note that, for both transactions, when the transaction send rate reaches above or close to the saturation point for maximum throughput, the latency increases significantly. This is due to the congestion caused by the rapid growth of the number of ordered transactions waiting in the validation queue, which affects the transaction committing latency; contributing to an increase in the average latency. In case of $Tx_{cr}$, as the new resource is being validated, the increased transaction send rate throttles the validator, increasing delays i.e. transaction endorsement latency. This affects the cumulative latency and throughput slows down. Figure~\ref{fig:createtl} shows that the throughput of creating new commodities decreases to minimum as the latency increases to its maximum.\par 
Thus, depending upon transaction send rate in a particular region of supply chain activity, we may need to distribute the transaction load among an appropriate set of validators.

\section{Conclusion}\label{sec:conc}
In  this  paper,  we  proposed  a  trust management framework for blockchain based supply chain applications to address the issue of trust associated with the quality of commodities and the entities logging data on the blockchain.  TrustChain design uses a consortium blockchain to track interactions among supply  chain  participants  and  to  dynamically  assign  trust  and reputation  scores  based  on  these  interactions. The framework also contributes in providing a reputation model which is both agent and asset based, it can assign  product-specific  reputations  for  the  same  participant, achieves automation and efficiency using smart contracts. We performed a qualitative security analysis with respect to threats in reputation systems. Performance analysis of a proof of concept implementation using Hyperledger showed that the  additional overhead introduced by Trustchain is minimal. 
In the future, we will explore how the different network models pertaining to consortium would affect the average throughput and latency of the system.

\ifCLASSOPTIONcaptionsoff
  \newpage
\fi

\bibliographystyle{IEEEtran}
\bibliography{mybib}

\begin{thebibliography}{10}
\providecommand{\url}[1]{#1}
\csname url@samestyle\endcsname
\providecommand{\newblock}{\relax}
\providecommand{\bibinfo}[2]{#2}
\providecommand{\BIBentrySTDinterwordspacing}{\spaceskip=0pt\relax}
\providecommand{\BIBentryALTinterwordstretchfactor}{4}
\providecommand{\BIBentryALTinterwordspacing}{\spaceskip=\fontdimen2\font plus
\BIBentryALTinterwordstretchfactor\fontdimen3\font minus
  \fontdimen4\font\relax}
\providecommand{\BIBforeignlanguage}[2]{{%
\expandafter\ifx\csname l@#1\endcsname\relax
\typeout{** WARNING: IEEEtran.bst: No hyphenation pattern has been}%
\typeout{** loaded for the language `#1'. Using the pattern for}%
\typeout{** the default language instead.}%
\else
\language=\csname l@#1\endcsname
\fi
#2}}
\providecommand{\BIBdecl}{\relax}
\BIBdecl

\bibitem{2013horse}
J.~Premanandh, ``Horse meat scandal--a wake-up call for regulatory
  authorities,'' \emph{Food control}, vol.~34, no.~2, pp. 568--569, 2013.

\bibitem{papaya}
\BIBentryALTinterwordspacing
(2017) Multistate outbreak of salmonella urbana infections linked to imported
  maradol papayas. [Online]. Available:
  \url{https://www.cdc.gov/salmonella/urbana-09-17/index.html}
\BIBentrySTDinterwordspacing

\bibitem{sidra2018productchain}
S.~{Malik}, S.~S. {Kanhere}, and R.~{Jurdak}, ``Productchain: Scalable
  blockchain framework to support provenance in supply chains,'' in \emph{2018
  IEEE 17th NCA Symposium}, Nov 2018, pp. 1--10.

\bibitem{IBMfraud}
\BIBentryALTinterwordspacing
(2019) Combating fraud with blockchain and crypto-anchors. [Online]. Available:
  \url{https://www.research.ibm.com/5-in-5/crypto-anchors-and-blockchain/}
\BIBentrySTDinterwordspacing

\bibitem{prov}
\BIBentryALTinterwordspacing
(2017) Provenance: Every product has a story. [Online]. Available:
  \url{https://www.provenance.org/}
\BIBentrySTDinterwordspacing

\bibitem{yang2017blockchain}
Z.~Yang, K.~Zheng, K.~Yang, and V.~C. Leung, ``A blockchain-based reputation
  system for data credibility assessment in vehicular networks,'' in \emph{2017
  IEEE 28th PIMRC Symposium}.\hskip 1em plus 0.5em minus 0.4em\relax IEEE,
  2017, pp. 1--5.

\bibitem{khaqqi2018incorporating}
K.~N. Khaqqi, J.~J. Sikorski, K.~Hadinoto, and M.~Kraft, ``Incorporating
  seller/buyer reputation-based system in blockchain-enabled emission trading
  application,'' \emph{Applied Energy}, vol. 209, pp. 8--19, 2018.

\bibitem{schaub2016trustless}
A.~Schaub, R.~Bazin, O.~Hasan, and L.~Brunie, ``A trustless privacy-preserving
  reputation system,'' in \emph{IFIP International Information Security and
  Privacy Conference}.\hskip 1em plus 0.5em minus 0.4em\relax Springer, 2016,
  pp. 398--411.

\bibitem{moinet2017blockchain}
A.~Moinet, B.~Darties, and J.-L. Baril, ``Blockchain based trust \&
  authentication for decentralized sensor networks,'' \emph{arXiv preprint
  arXiv:1706.01730}, 2017.

\bibitem{cachin2016architecture}
C.~Cachin, ``Architecture of the hyperledger blockchain fabric,'' in
  \emph{Workshop on Distributed Cryptocurrencies and Consensus Ledgers}, vol.
  310, 2016.

\bibitem{starbird1997acceptance}
S.~A. Starbird, ``Acceptance sampling, imperfect production, and the optimality
  of zero defects,'' \emph{Naval Research Logistics (NRL)}, vol.~44, no.~6, pp.
  515--530, 1997.

\bibitem{fsanz}
\BIBentryALTinterwordspacing
(2019) Food standards australia new zealand. [Online]. Available:
  \url{http://www.foodstandards.gov.au/Pages/default.aspx}
\BIBentrySTDinterwordspacing

\bibitem{commerce2002beta}
B.~E. Commerce, A.~J{\o}sang, and R.~Ismail, ``The beta reputation system,'' in
  \emph{In Proceedings of the 15th Bled Electronic Commerce Conference}.\hskip
  1em plus 0.5em minus 0.4em\relax Citeseer, 2002.

\bibitem{gambetta2000can}
D.~Gambetta \emph{et~al.}, ``Can we trust trust,'' \emph{Trust: Making and
  breaking cooperative relations}, vol.~13, pp. 213--237, 2000.

\bibitem{etsi2011102}
T.~ETSI, ``102 165-1 v4. 2.3 (2011-03),'' \emph{Technical Specification
  Telecommunications and Internet converged Services and Protocols for Advanced
  Networking (TISPAN)}, p.~75, 2011.

\bibitem{diallo2018trustworthy}
M.~H. Diallo, N.~Panwar, S.~Mehrotra, and A.~A. Sani, ``Trustworthy sensing in
  an untrusted iot environment,'' in \emph{2018 IEEE PerCom Workshops}, 2018,
  pp. 468--471.

\bibitem{brodsky2018overlapping}
W.~L. Brodsky, J.~R. Dangler, P.~D. Isaacs, D.~C. Long, and M.~T. Peets,
  ``Overlapping, discrete tamper-respondent sensors,'' Mar.~6 2018, uS Patent
  9,911,012.

\bibitem{saberi2018blockchain}
S.~Saberi, M.~Kouhizadeh, J.~Sarkis, and L.~Shen, ``Blockchain technology and
  its relationships to sustainable supply chain management,''
  \emph{International Journal of Production Research}, pp. 1--19, 2018.

\bibitem{thakkar2018performance}
P.~Thakkar, S.~Nathan, and B.~Viswanathan, ``Performance benchmarking and
  optimizing hyperledger fabric blockchain platform,'' in \emph{2018 IEEE 26th
  MASCOTS Symposium}, 2018, pp. 264--276.

\end{thebibliography}

\end{document}